\documentclass[a4paper]{JAC2003}
\usepackage{graphicx}
\usepackage{booktabs}
\usepackage{amsmath}

\usepackage{multirow}
\usepackage{subfigure}

\usepackage{varwidth}
\usepackage{xcolor}

\begin{document}
\title{BEAM--BEAM EFFECTS IN BEPCII}

\author{Y. Zhang\\ Institute of High  Energy Physics, Beijing, China}

\maketitle

\begin{abstract}
  We first introduce the design parameters of the Beijing Electron--Positron Collider II (BEPCII) and the simulation
  study of beam--beam effects during the design process of the machine.
  The main advances since 2007 are briefly introduced and reviewed. The
  longitudinal feedback system was installed to suppress the coupled
  bunch instability in January 2010. The horizontal tune decreased
  from 6.53 to 6.508 during the course of data taken in December,
  2010. The saturation of the beam--beam parameter was found in 2011, and
  the vacuum chambers and magnets near the north crossing point were moved
  15~cm in order to mitigate the long range beam--beam interaction. At
  the beginning of 2013, the beam--beam parameter achieved 0.04 with the
  new lower $\alpha_p$ lattice and the peak luminosity achieved
  $7\times 10^{32}\text{~cm}^{-2}\text{~s}^{-1}$.
\end{abstract}

\section{Introduction}
The Beijing Electron--Positron Collider (BEPC) was constructed for both
high energy physics and Synchrotron Radiation (SR) research. As a unique
$e^+e^-$ collider operating in the $\tau$-charm region and the
first SR source in China, the machine has been operated for well over
16 years since it was put into operation in 1989.

BEPCII is an upgrade project from BEPC. It is a double ring machine.
Following the success of KEKB, the crossing scheme was adopted in
BEPCII, where two beams collide with a horizontal crossing angle of
$2\times 11$~mrad. The design luminosity of BEPCII is $1.0\times
10^{33}\mbox{~cm}^{-2}\mbox{~s}^{-1}$ at $1.89~$GeV, which is about $100$
times higher than BEPC~\cite{bepcii-report}. The main design collision
parameters are shown in Table~\ref{tab:design}.
\begin{table}[hbt]
  \centering
  \caption{Design Parameters of BEPCII.}
  \label{tab:design}
  \begin{tabular}{l|l||l|l}
    \toprule
      E & 1.89 GeV &       $\nu_s$ & 0.034 \\
      $I$ & 910 mA &   $\alpha_p$ & 0.024 \\
      $I_b$ & 9.8 mA &  $\sigma_{z0}$ & 0.0135 m\\
      $n_b$ & 93 &       $\sigma_z$ & 0.015 m \\
      $V_{rf}$ & 1.5 MV  &     $\epsilon_x$ & 144 nmrad \\
      $\beta_x^*/\beta_y^*$ & 1.0/0.015 m &  Coupling & 1.5\% \\
      $\nu_x/\nu_y$ & 6.53/5.58 &   $\xi_y$ & 0.04 \\
      $\theta_c$ & 22 mrad &
      $\tau_x/\tau_y/\tau_z$ & 3.0e4/3.0e4/1.5e4\\
      \bottomrule
    \end{tabular}
\end{table}
In March 2013, the peak luminosity achieved $7.0\times
10^{32}\text{~cm}^{-2}\text{~s}^{-1}$ with 120 bunches and beam current
730~mA, where a lower $\alpha_p$ lattice was used.

In the following, we first introduce the simulation study of the beam--beam
interaction during the design course of the machine. Then we review
the performance and optimization of the real machine.

\section{Simulation Study During Design}

\subsection{Code Development}
We have developed new parallel strong--strong beam--beam code, which
is used to study the effects in BEPCII~\cite{zhangy-prst05}. The main characteristics of
the code are listed below.
\begin{itemize}
\item Particle-in-cell. The Triangular Shaped Cloud (TSC) method is
  employed for the charge assignment, where the charge of each
  macroparticle is assigned to its nine nearest points by weight.
\item Synchrotron motion is included. The transportation through the
  arc is same as that of Hirata's BBC code~\cite{hirata-bbc}.
\item The beam--beam potential is calculated by solving the Poisson
  equation with open boundary.
\item Bunch length effect is included by longitudinal slices and the
  interpolation of beam--beam potential is done when the collision
  between two slices is considered,  which helps to reduce the required
  slice number.
\item It is assumed that a particle in one slice will not jump into
  non-adjacent ones on the next turn. It seems that this assumption has been valid so far, especially in the ordinary collision scheme (where
  the required slice number is only about 5).
\item Lorentz boost is used to consider the crossing angle effect~\cite{hirata-crossing}.
\end{itemize}

\subsection{Code Check}

\begin{itemize}
\item The geometrical factor of luminosity reduction for head-on
  collision. The loss in luminosity due to geometrical effect for
  nominal BEPCII parameters is 86\%, and the code tracking result
  agrees well.
\item The geometrical factor of luminosity reduction for collision
  with finite crossing angle. The loss in luminosity due to geometrical
  effect for design BEPCII parameters is 80\%, and the code tracking
  result agrees well.
\item The beam--beam field calculated by the code for the Gaussian beam
  distribution agrees well with the Bassetti--Erskine formula.
\item The synchro-betatron mode agrees well with that predicted by the hollow beam
  matrix model~\cite{hollow-matrix}.
\item The luminosity result for BEPCII agrees well with that of K. Ohmi's
  code~\cite{ohmi-code}.
\end{itemize}

\subsection{Simulation Result}

The achieved beam--beam parameter $\xi$ with collision is defined as
\begin{equation}
  \xi_u = \frac {N r_e} {2\pi \gamma} \frac {\beta_u^0} {\sigma_u (\sigma_x +
    \sigma_y)}
\end{equation}
where $\beta^0$ is the nominal beta function without collision, and
$\sigma$ is the disturbed beam size with collision. If we don't consider
the finite bunch length and finite crossing angle, the bunch
luminosity can be represented as
\begin{equation}
  L = \frac {N^2 f_0} {4\pi \sigma_x \sigma_y}
\end{equation}
where $\sigma$ is the disturbed beam size with collision. In the normal
case, $\sigma_y \ll \sigma_x$, the achieved $\xi_y$ can be represented by luminosity,
\begin{equation}
  \xi_y = \frac {2r_e\beta_y^0}{N\gamma} \frac{L}{f_0}.
\end{equation}

\begin{figure}[htb]
  \centering
  \includegraphics*[width=65mm]{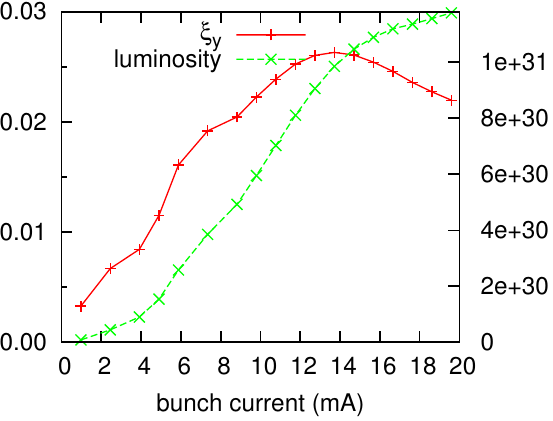}
  \caption{The achieved $\xi_y$ and bunch luminosity versus bunch
    current with design parameters.}
  \label{fig:limit-sim-x53}
\end{figure}

With the design parameters, the maximum $\xi_y$ only achieves 0.025,
which is shown in Fig.~\ref{fig:limit-sim-x53}. This is bad news for
the BEPCII team, since $\xi_y$ needs to achieve 0.04 if we want to achieve the
designed luminosity with the designed beam current. We therefore did some
estimation to determine if it would be feasible to inject more bunches, and it
seems that this would be possible.

\begin{figure}
  \centering
  \includegraphics*[width=65mm]{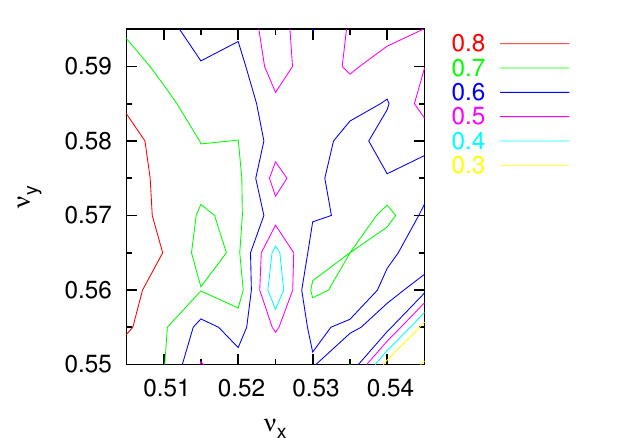}
  \caption{Tune survey of luminosity. The luminosity is normalized to
    the design value.}
  \label{fig:tunescan}
\end{figure}

The beam--beam performance is very sensitive to the working point. The
normalized luminosity versus tune is depicted in
Fig.~\ref{fig:tunescan}. The best working point is near (0.505,0.570),
where the luminosity is about 80\% of the design value. That is to say, we
could achieve $8\times 10^{32}\text{~cm}^{-2}\text{~s}^{-1}$ with the designed
bunch current, bunch number and optimized working point.

The full horizontal crossing angle between colliding beams is
22~mrad. The luminosity reduction factor is less than 10\% at
(0.53,0.58), however it is about 30\% at (0.51,0.57). It seems that the
luminosity loss due to a finite crossing angle is more serious the closer the
horizontal tune is to $0.5$, the high luminosity
working point region. 

We also tried to analyze the coupling contribution and carried out some simulations at
different working points. The results are summarized in
Table~\ref{tab:coupling}. It seems that we have to move the horizontal
tune closer to $0.5$ and ensure that the emittance coupling is
less than 0.5\% if $\xi_y$ is expected to achieve 0.04. 
\begin{table}
  \centering
  \caption{Coupling contribution at different working point.}
  \label{tab:coupling}
  \begin{tabular}{l|l|l|l}
    \toprule
    Tune & Coupling  & Max $\xi_y$ & Lum  \\
    \hline
    \multirow{3}{*}{(0.510, 0.575)}
      & 0.5\% & 0.041@11~mA & 12.3e30 \\
      & 1.0\% & 0.037@12~mA & 12.1e30 \\
      & 1.5\% & 0.034@13~mA & 12.1e30 \\
      \hline
      \multirow{2}{*}{(0.530, 0.580)}
      & 0.5\% & 0.026@7~mA & 5.0e30  \\
      & 1.5\% & 0.026@13~mA & 9.2e30  \\
      \hline
      \multirow{3}{*}{(0.535, 0.575)}
      & 0.5\% & 0.031@9~mA & 7.6e30  \\
      & 1.0\% & 0.027@9~mA & 6.6e30  \\
      & 1.5\% & 0.023@9~mA & 5.6e30  \\
      \hline
      \multirow{2}{*}{(0.540, 0.590)}
      & 0.5\% & 0.025@11~mA & 7.6e30  \\
      & 1.0\% & 0.024@11~mA & 7.2e30  \\
      \bottomrule
    \end{tabular}
  \end{table}

\section{Performance and Optimization}

The first electron beam was stored in the SR ring in November
2006. Optics measurement and correction was studied at that time. The
backup collision mode was first tuned in the spring of 2007, during the course of which we
learned the collision tuning. The superconducting
final focus magnet was installed in the summer of 2007. The detector
was installed in June 2008, and this completed the construction of the
machine. Here, we review the machine tuning history
in chronological order.

\subsection{Phase I: Autumn of 2008 to Summer of 2010}

The big events in this period are listed below.
\begin{itemize}
\item January 2009. Profile monitor, which caused very strong longitudinal multibunch
  instability, was removed from the positron
  ring. 
\item May 2009. Horizontal tune was moved to $0.51$ from
  $0.53$. Luminosity reached $3\times
  10^{32}\text{~cm}^{-2}\text{~s}^{-1}$, which is the `design goal' of the
  government funding agency.
\item January 2010. Longitudinal feedback system was installed and
  began to work.
\end{itemize}

\begin{figure}
  \centering
  \includegraphics*[width=65mm]{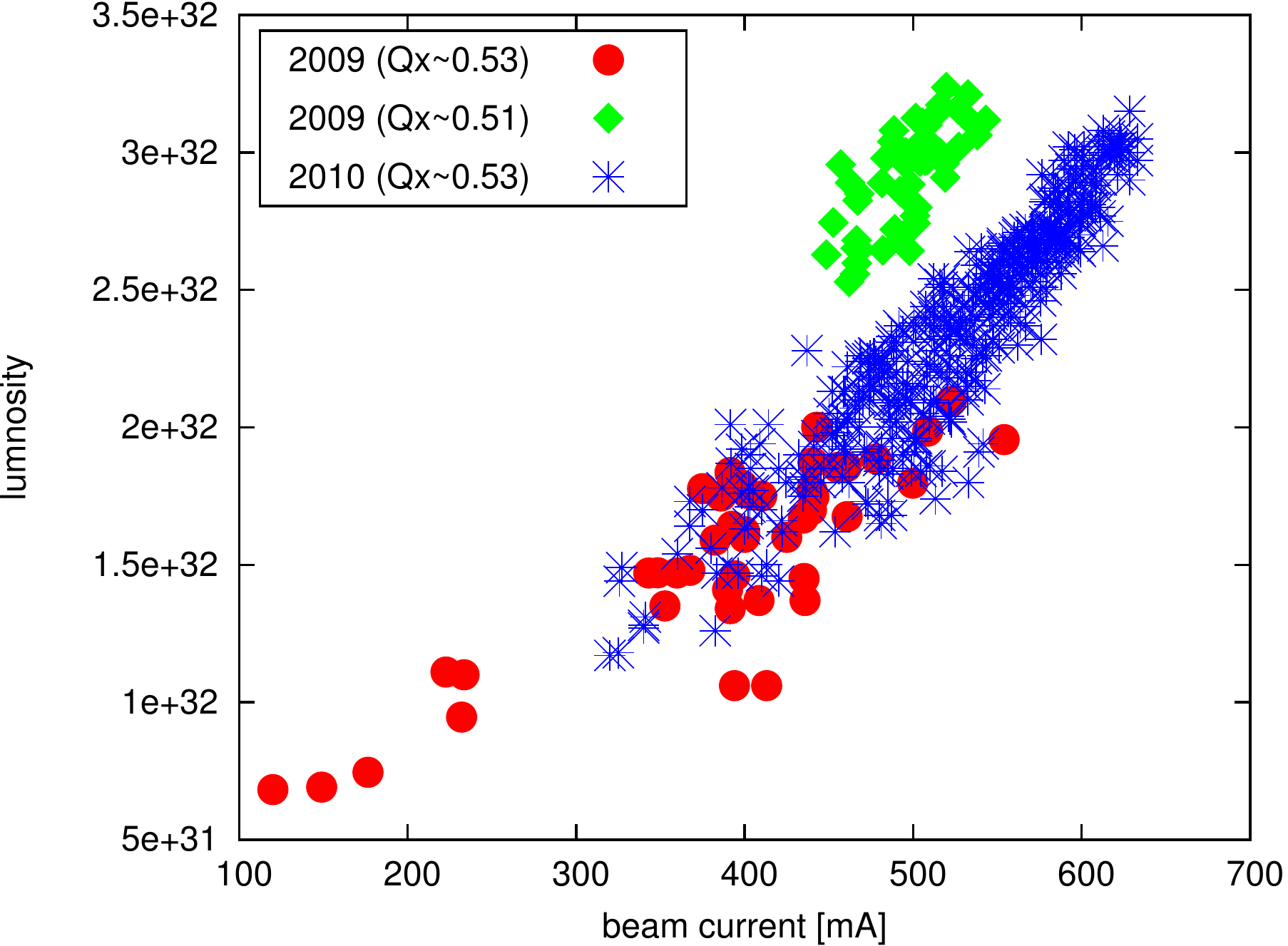}
  \caption{Luminosity versus beam current in 2009 and 2010. The
    difference between the red (2009) and blue (2010) dots comes from
    the suppression of longitudinal multibunch instability.}
  \label{fig:lum-2009-2010}
\end{figure}

\begin{figure}
  \centering
  \subfigure[]{\includegraphics*[width=40mm]{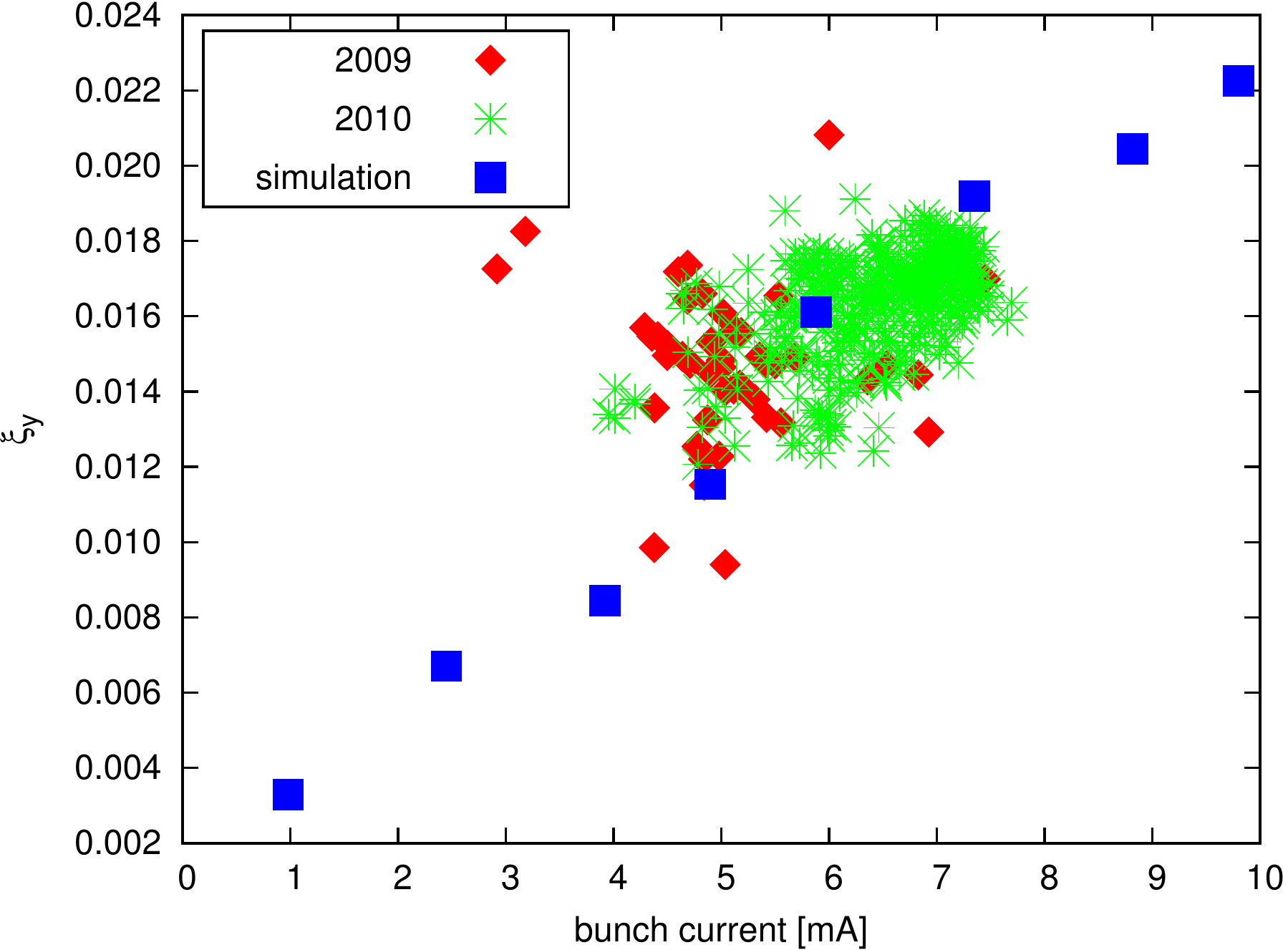}}
  \subfigure[]{\includegraphics*[width=40mm,height=30mm]{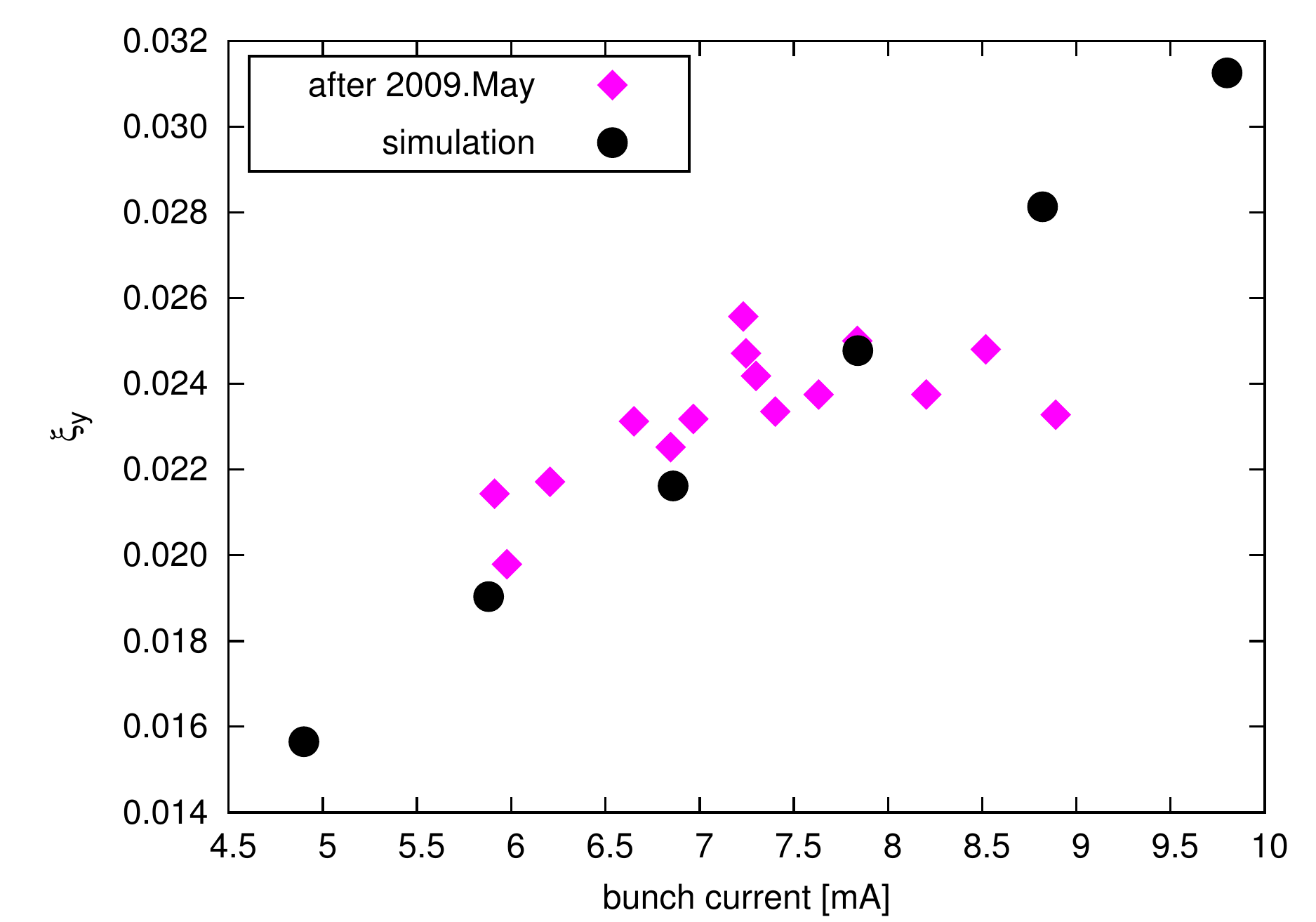}}
  \caption{Beam--beam parameter versus bunch current. The left figure
    shows that $Q_x\sim 0.53$ and the right shows that $Q_x\sim
    0.51$. The longitudinal feedback system begins to work in 2010.}
  \label{fig:xi-2009-2010}
\end{figure}

Figure~\ref{fig:lum-2009-2010} shows the luminosity versus beam current
and Fig.~\ref{fig:xi-2009-2010} shows the beam--beam parameter versus
bunch current. The longitudinal coupled bunch instability still reduced
the luminosity performance even after the removal of the profile monitor, which caused  very strong instability,
from the positron ring.  In
order to increase the luminosity with the same beam current, we tried to
move the horizontal tune closer to $0.5$ in May 2009. The peak
luminosity increased from 2 to $3\times
10^{32}\text{~cm}^{-2}\text{~s}^{-1}$. Since the detector background is
too high to take data with $Q_x\sim 0.51$, the machine continued to run
with $Q_x\sim 0.53$ in the following normal data collection run. In the first half of 2010, the longitudinal feedback
system began to work and the peak luminosity achieved $3\times
10^{32}\text{~cm}^{-2}\text{~s}^{-1}$ with $Q_x\sim 0.53$. The maximum
$\xi_y$ is about 0.02 when $Q_x\sim 0.53$, which is less than the
simulated 30\% percent (see Table~\ref{tab:coupling}).

\subsection{Phase II: Autumn of 2010 to Summer of 2011}

The big events in this period are listed below.
  \begin{itemize}
  \item July 2010. It was found that the final focus magnet and
    vacuum chamber on one side of the detector was displaced by about
    10~mm in the horizontal direction. It was aligned in the summer
    shutdown.
  \item December 2010. Detector background was reduced with $Q_x\sim
    0.51$. The physics people could take data near the 0.51 working
    point.
  \end{itemize}

\begin{figure}
  \centering
  \includegraphics*[width=65mm]{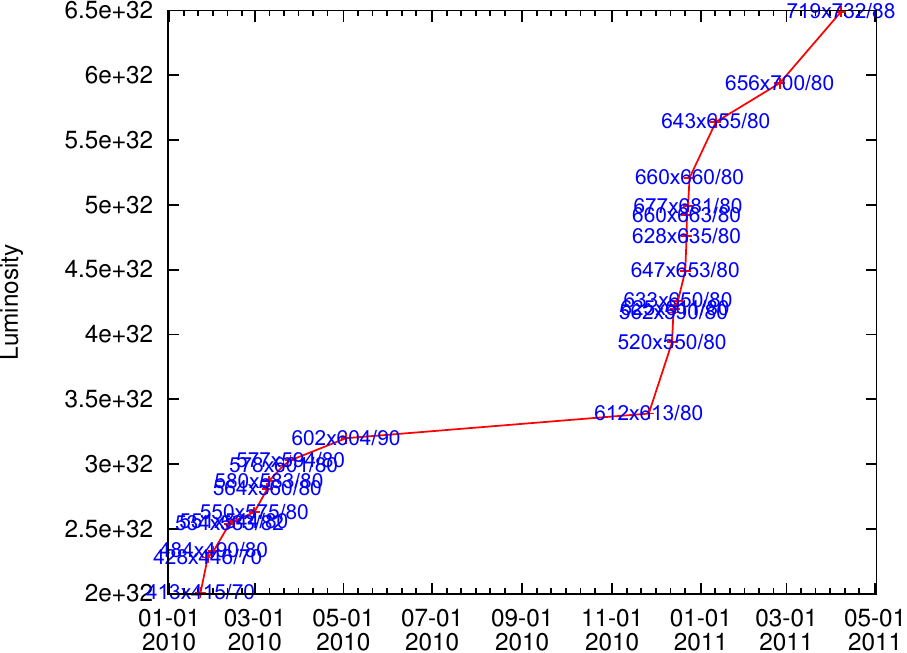}
  \caption{Peak luminosity record from the beginning of 2010 to the
    summer of 2011. The colliding beam current and bunch number is
    also shown in the figure.}
  \label{fig:peaklum-2011}
\end{figure}

\begin{figure}
  \centering
  \includegraphics*[width=65mm]{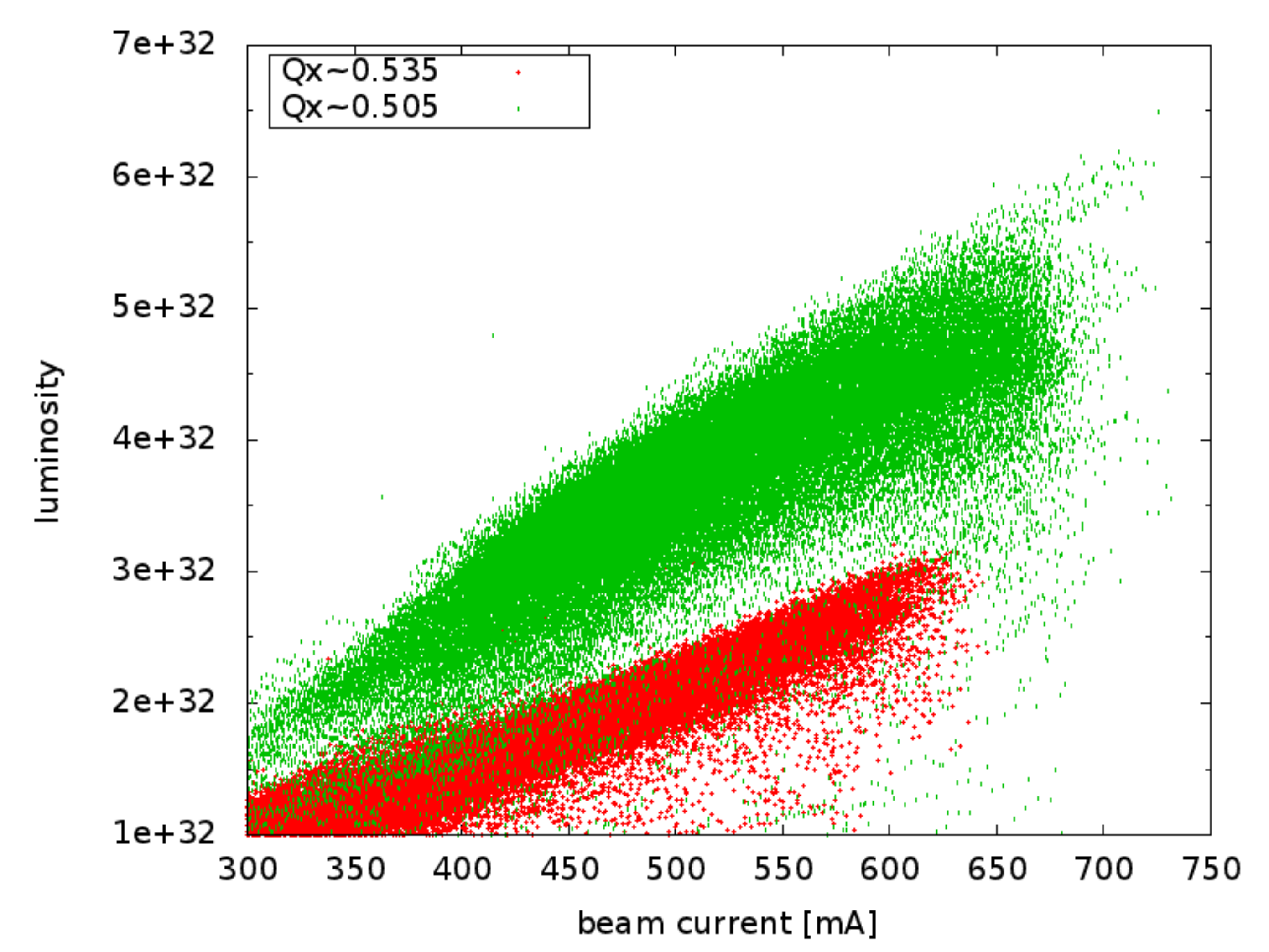} 
  \caption{Luminosity performance at different working points.}
  \label{fig:lum-comp-tune}
\end{figure}

  The most important advance in this period is the reduction of
  the detector background with $Q_x \sim 0.51$, since the physics people
  could take data at the working point and the accelerator people had
  enough time to do the luminosity tuning. The detector background is
  mainly optimized by the closed orbit tuning along the
  ring. Figure~\ref{fig:peaklum-2011} shows the peak luminosity record
  from the beginning of 2010 to the summer of 2011. It was very
  exciting near the start of 2011 since a new record would be born
  only in a few days. The peak luminosity was $6.5\times 10^{32}
  \text{~cm}^{-2}\text{~s}^{-1}$ in 2011. The comparison of luminosity
  at different working points is shown in Fig.~\ref{fig:lum-comp-tune},
  which very obviously shows that a working point closer to $0.5$ means a higher
  luminosity. 

\begin{figure}
  \centering
  \includegraphics*[width=65mm]{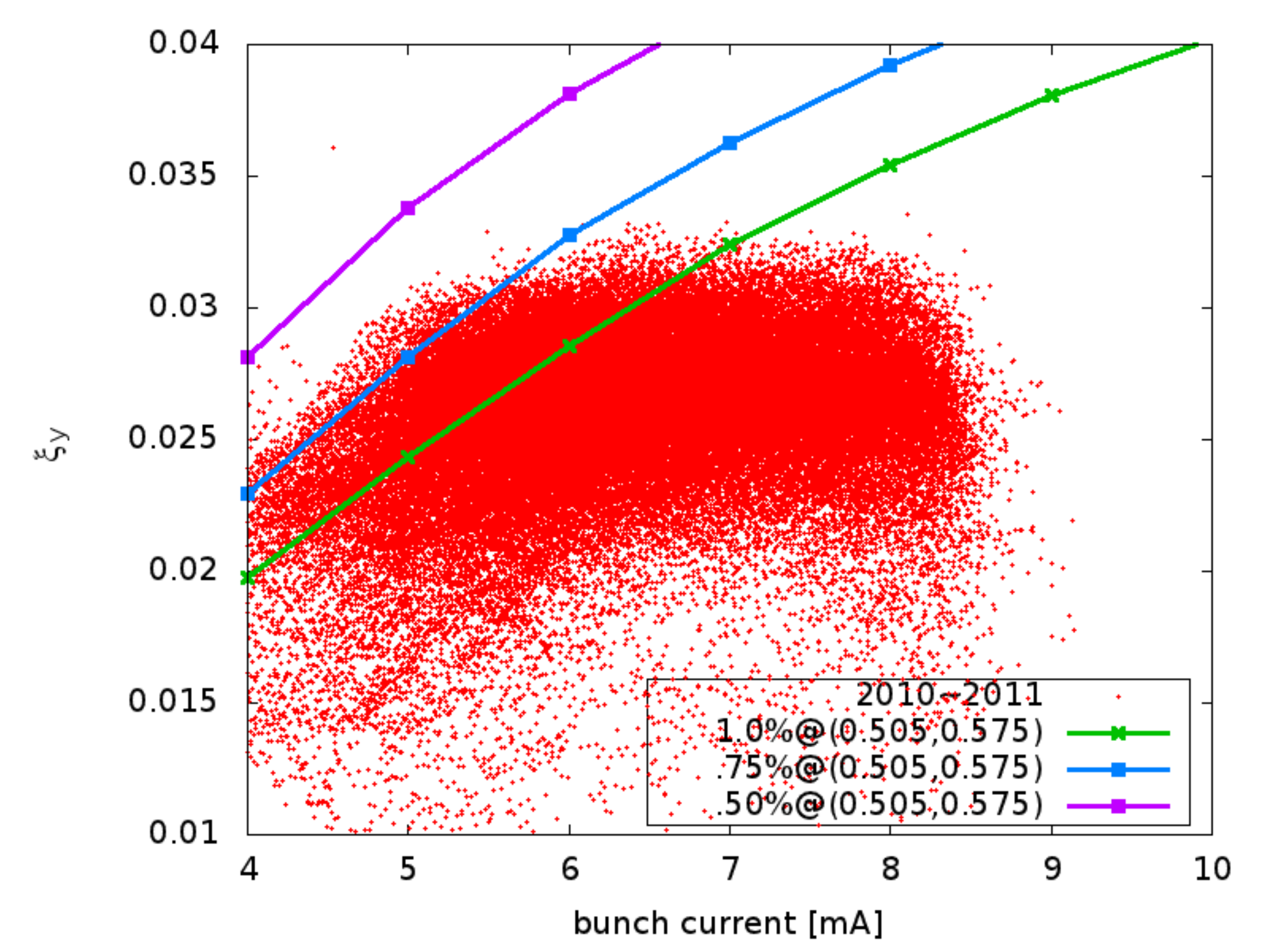} 
  \caption{Achieved beam--beam parameter in 2010--2011. The
    lines shows the simulation result with different coupling.}
  \label{fig:xiy-2011}
\end{figure}

The 2010--2011 commissioning year was very successful and exciting, but there was
some confusion  when we saw the beam--beam performance. The achieved
$\xi_y$ is shown in Fig.~\ref{fig:xiy-2011}. There exist clear
saturation phenomenon for $\xi_y$ and the maximum is about 0.033. We
should explain what caused the saturation.

\begin{figure}
  \centering
  \includegraphics*[width=65mm]{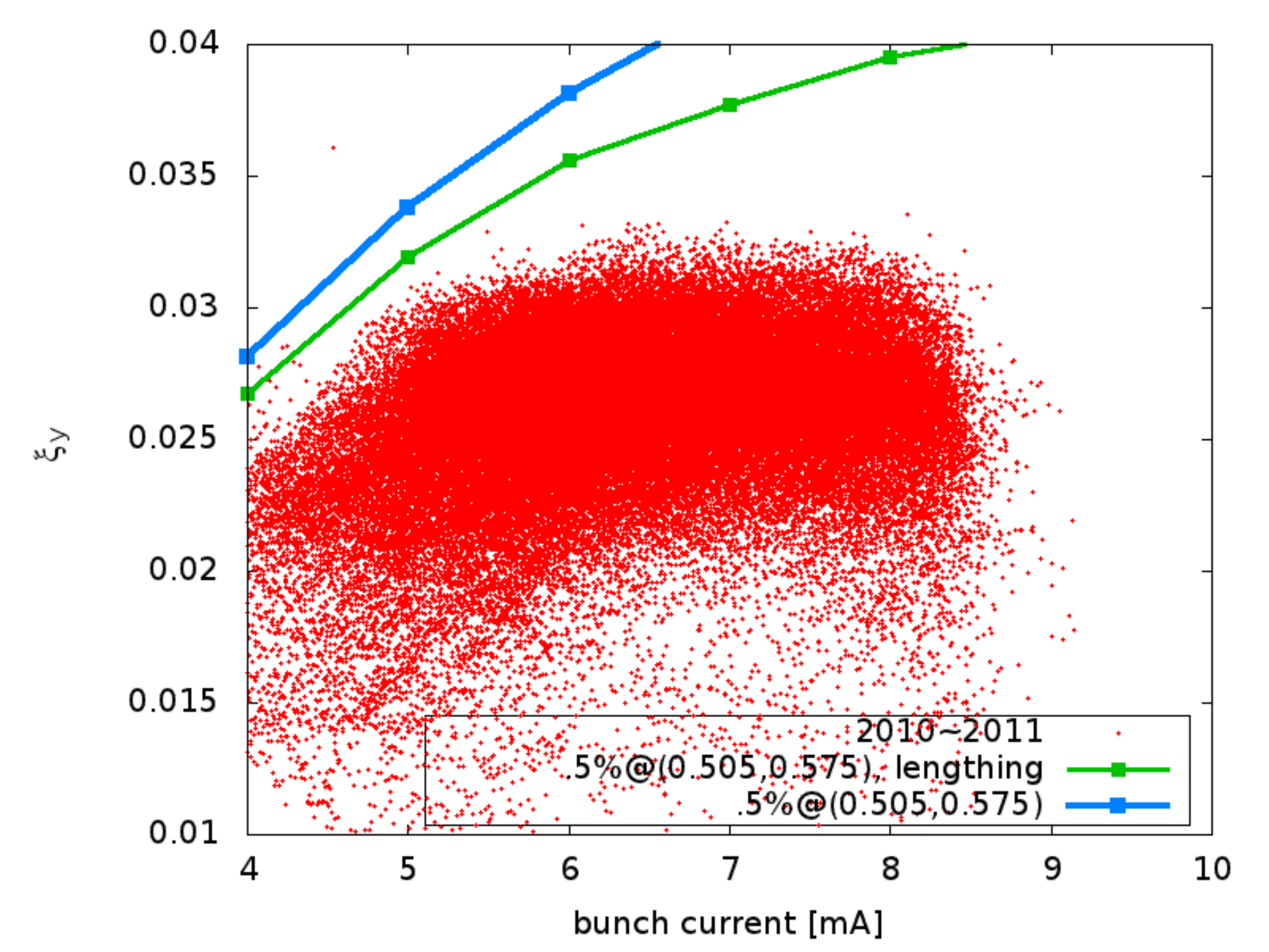} 
  \caption{Beam--beam loss due to bunch lengthening, which does not
    explain the beam--beam parameter saturation in real machine.}
  \label{fig:xiy-lengthing-2011}
\end{figure}

Figure~\ref{fig:xiy-lengthing-2011} shows the bunch lengthening effect. It
seems this effect does not bring very much luminosity loss, and the
maximum beam--beam parameter is still above 0.04.

\begin{figure}
  \centering
  \includegraphics*[width=65mm]{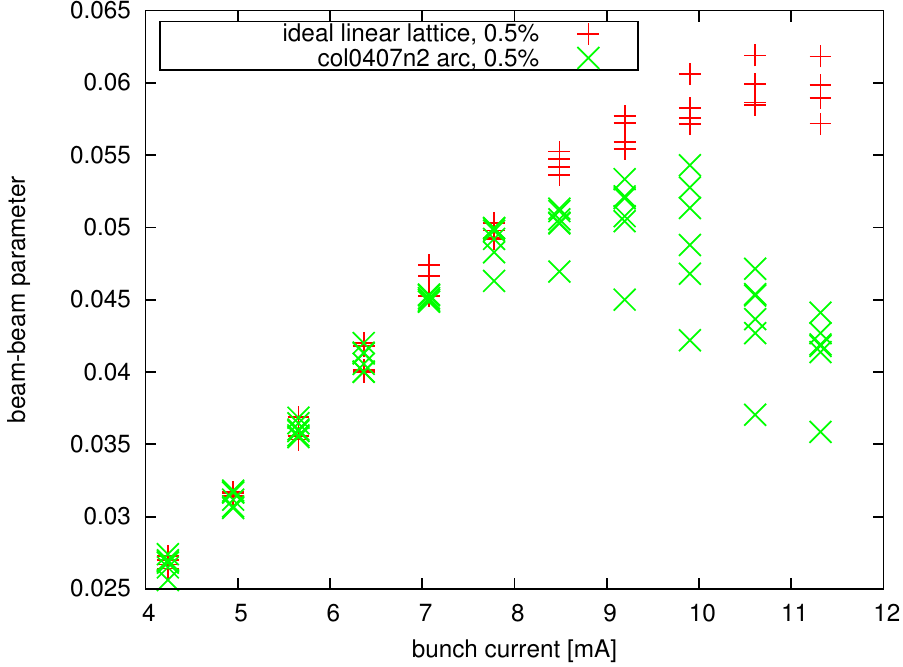}
  \caption{The luminosity loss due to the realistic arc. The arc map
    is implemented using element-by-element tracking. The ideal linear
  lattice  means the map is represented by a transfer matrix.}
  \label{fig:xiy-arc}
\end{figure}

The nonlinear arc may also reduce the luminosity performance. We use
Hirata's BBC~\cite{hirata-bbc} code as a pass method in Accelerator
Toolbox (AT)~\cite{at} to simulate the weak--strong beam--beam
interaction. The map in the arc is implemented using the
element-by-element symplectic tracking in AT. Figure~\ref{fig:xiy-arc}
shows the comparison between the ideal transfer matrix map and
element-by-element tracking in arc. The lattice really reduces the
beam--beam performance, but we did not believe that the saturation was mainly
caused by the crosstalk between nonlinear arc and beam--beam force. On
the other hand, we could not ignore the simulation result, which told
us that we should put more emphasis on the sexupole optimization.

\begin{figure}
  \centering
  \includegraphics*[width=65mm]{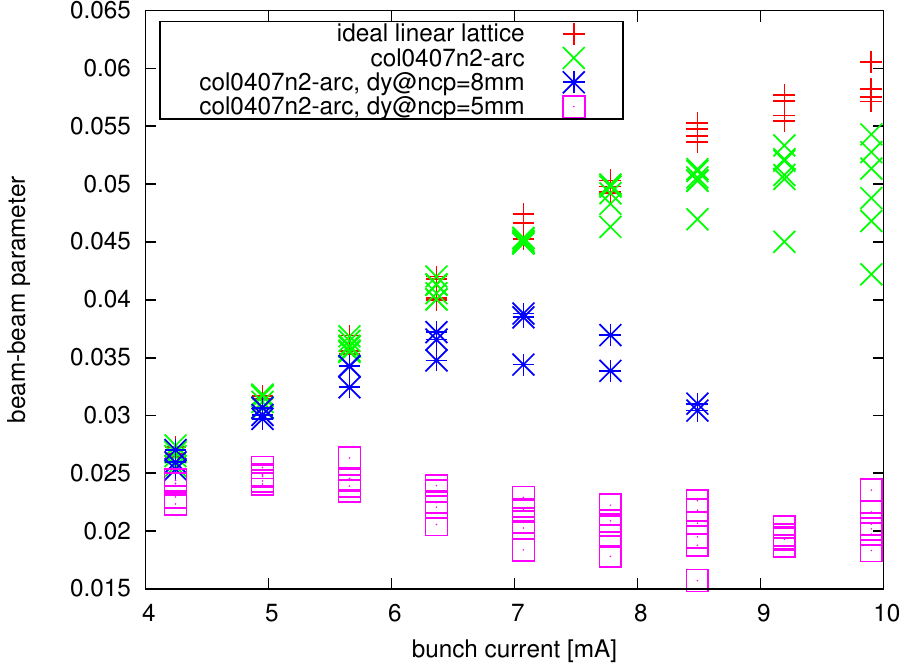}
  \caption{The luminosity loss due to nonlinear arc and long range
    beam--beam interaction at NCP.}
  \label{fig:xiy-ncp}
\end{figure}

There is another crossing point (NCP) in the north of the two rings,
where the colliding beams are separated vertically by about 8~mm and the
full horizontal angle is about $2\times 0.15$~rad (17.7$^\circ$). We
still use the weak--strong code (AT and BBC) to study the parasitic
beam--beam effect, which is shown in Fig.~\ref{fig:xiy-ncp}. The
achieved $\xi_y$ is only about 0.035 with 8~mm separation at NCP

\subsection{Phase III: Autumn of 2011 to Summer of 2012}

The big events in this period are listed below.
\begin{itemize}
\item In the summer shutdown of 2011, the NCP chambers and magnets
  were moved 15 cm, 1/4 of the rf bucket. The horizontal separation
  between colliding bunches is greater than $30\sigma_x$.
\end{itemize}

\begin{figure}
  \centering
  \includegraphics*[width=65mm]{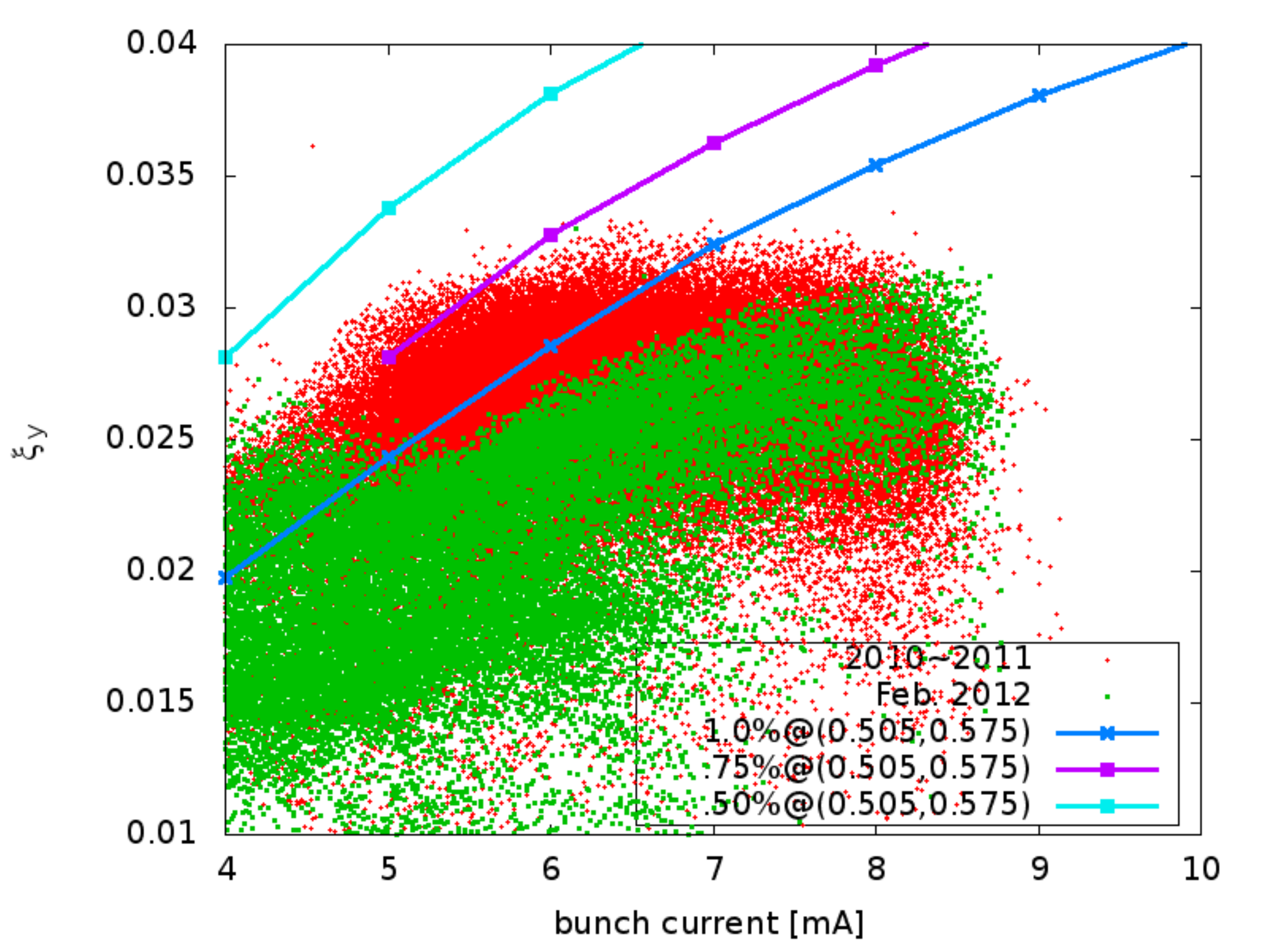} 
  \caption{Beam--beam parameter achieved in 2011 and 2012. The
    mitigation of long range beam--beam interaction at NCP did not
    bring any improvements.}
  \label{fig:xiy-2012}
\end{figure}

After the hardware modification, the beam--beam performance did not
increase as expected, which is shown in Fig.~\ref{fig:xiy-2012}. This
could be explained to some extent by the large longitudinal offset of
the collision point. In 2011--2012 commissioning year, the offset is about
3~mm, and it is about 6~mm in February 2012. We did not put enough
emphasis on monitoring the parameter during collision at that
time.

\subsection{Phase IV: Autumn of 2012 to March 2013}

The big events in this period are listed below.
\begin{itemize}
\item Lower $\alpha_p$  mode was first tested at 2.18~GeV in February
  2013. The $\xi_y$ record 0.033 was broken after about two years.
\item One bunch every three buckets, and even one bunch every two
  buckets, injection was tested in the machine study of March 2013. The
  peak luminosity achieved $7.0\times
  10^{32}\text{~cm}^{-2}\text{~s}^{-1}$ at 1.89~GeV.
\end{itemize}

The momentum compaction factor of the new lattice is about 0.017, and the
old one is 0.024. The reduction of $\alpha_p$ is achieved by increasing
the horizontal tune from 6.5 to 7.5. During the lattice design, we
also optimized the chromatic distortion and some nonlinear resonance driving terms. 
However we still did not establish a so-called
`standard' that could tell us if the lattice is good enough.

\begin{figure}
  \centering
  \includegraphics*[width=65mm]{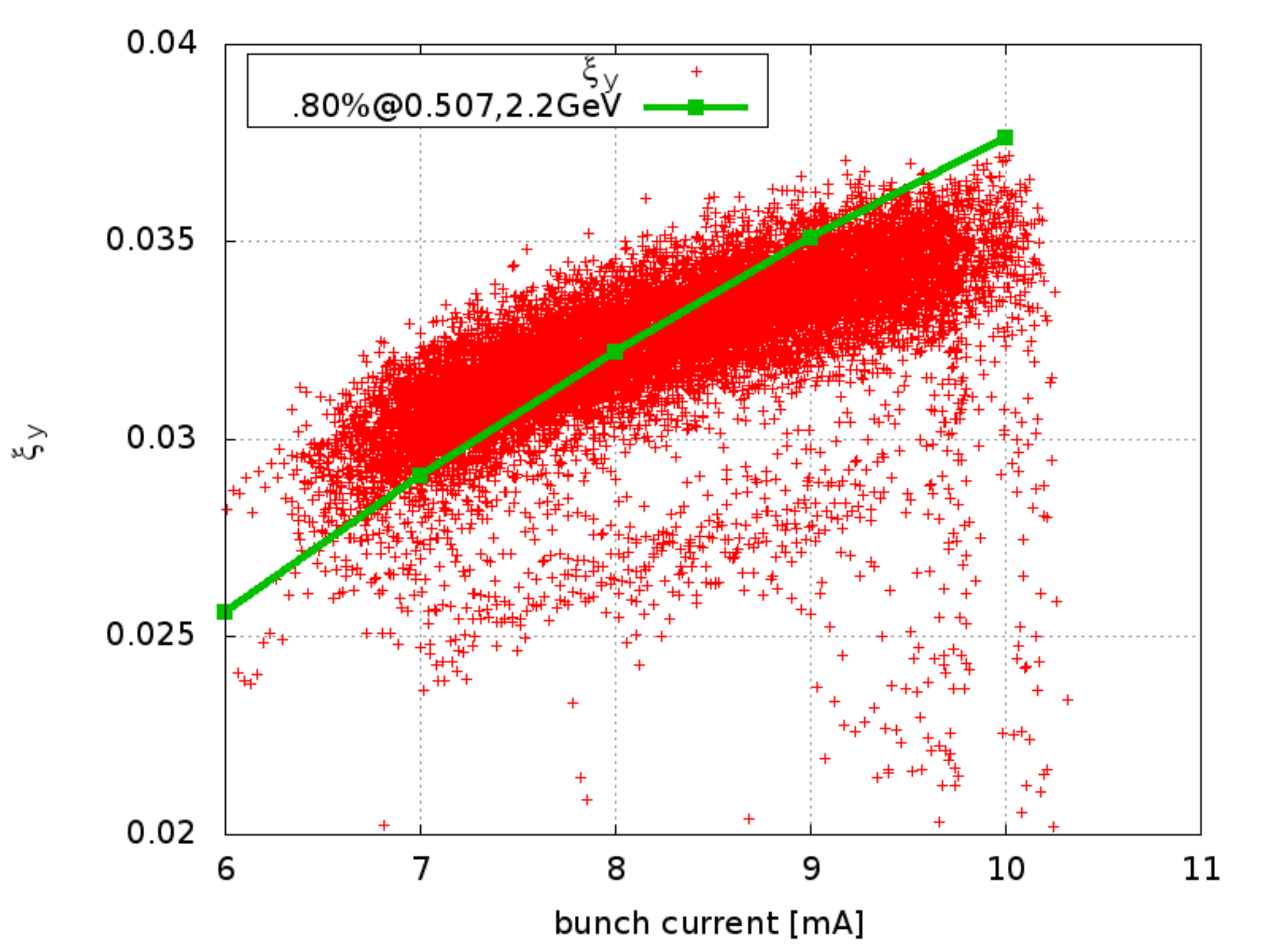} 
  \caption{Achieved beam--beam parameter at 2.18~GeV with new lattice in
    2013.}
  \label{fig:xiy-alphap-2.18}
\end{figure}

\begin{figure}
  \centering
  \includegraphics*[width=65mm]{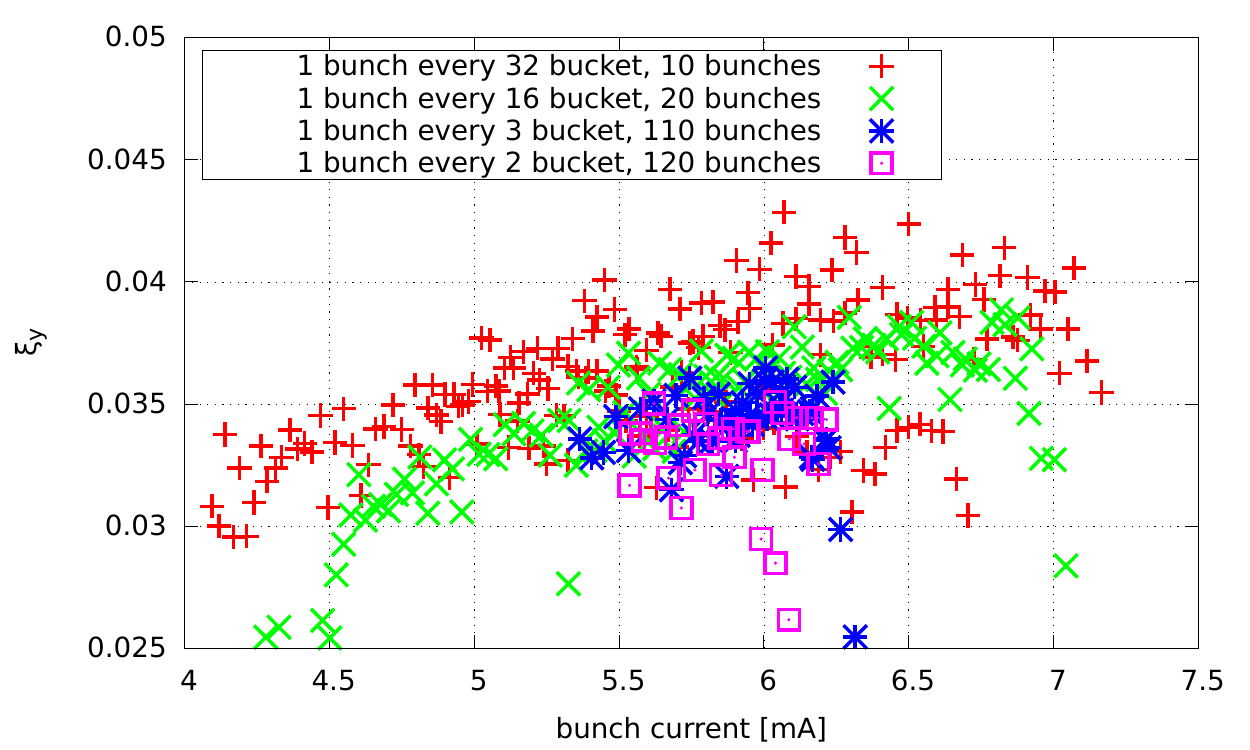}
  \caption{Achieved beam--beam parameter at 1.89~GeV with different
    bunch pattern in 2013.}
  \label{fig:xiy-md189-2013}
\end{figure}

The achieved beam--beam performance at 2.18~GeV is shown in
Fig.~\ref{fig:xiy-alphap-2.18}. We also did some machine study in order
to increase the peak luminosity at 1.89~GeV. The achieved beam--beam
parameter with different bunch pattern is shown in
Fig.~\ref{fig:xiy-md189-2013}. The maximum $\xi_y$ is above 0.04. It
seems that the multibunch effect reduces the beam--beam performance, which
would be a serious limitation if we were to continue to increase the luminosity.

\section{Summary}

We review the collision optimization history of BEPCII. The suppression of multibunch longitudinal instability and
moving the horizontal tune close to $0.5$ helped us to
increase the luminosity. The mitigation of long range beam--beam
interaction seems not so effective as expected, indicating that maybe the real
vertical separation is greater than estimated. The lower $\alpha_p$
lattice helped us to achieve the $\xi_y$ record of 0.04 at 1.89~GeV.

The simulation study is very important both in the design and the daily
commissioning. It gives a benchmark in normal operation and lets us
know if the status is optimized enough, even though we could approach
the simulation result and never go beyond it. The difference between the
simulation and the optimized result is about 10--20\%. It should also be
emphasized that we would like to use the maximum achieved $\xi_y$ in
the simulation as the beam--beam limit in the simulation.

Increasing beam current is a must to increasing the luminosity. However,
it seems the multibunch effect is very serious. The study to cure the
instability and even find the instability source will be very important in
the future. In the near future, we'll test a new lattice with
$alpha_p$ about 0.017, larger emittance (100~nm$\rightarrow$130~nm) and lower
$\beta_y$ (1.5~cm$\rightarrow$1.35~cm). The colliding bunch current could be higher
with the new mode and the beam current could be higher with same bunch
number. It is expected that this could help us to increase the luminosity.


\begin{thebibliography}{9}   

\bibitem{bepcii-report}
  Design Report of BEPCII--Accelerator Part (2nd ed) (2003).

\bibitem{zhangy-prst05}
  Y. Zhang, K. Ohmi and L. Chen, Phys. Rev. ST Accel. Beams 8 (2005) 074402.

\bibitem{hirata-bbc}
  BBC: Program for Beam--Beam Interaction with Crossing Angle. http://wwwslap.cern.ch/collective/hirata/

\bibitem{hirata-crossing}
  K. Hirata, Phys. Rev. Lett. 74 (1995) 2228-2231.

\bibitem{hollow-matrix}
  E.G. Stern, J.F. Amundson, P.G. Spentzouris, and A.A. Valishev,
  Phys. Rev. ST Accel. Beams 13 (2010) 024401.

\bibitem{ohmi-code}
  K. Ohmi, Phys. Rev. E 62 (2000) 7287.
\bibitem{at}
  A.Terebilo, SLAC-PUB-8732 (2001).


\end{thebibliography}
\end{document}